\documentclass[runningheads]{llncs}
\usepackage{graphicx}
\usepackage{cite}
\usepackage{booktabs}
\usepackage{makecell}
\usepackage{longtable}
\usepackage{multirow}

\begin{document}
\title{How to characterize the health of an Open Source Software project? A snowball literature review of an emerging practice}
\titlerunning{How to characterize the health of an Open Source Software project?}
%
\author{Johan Linåker\inst{1}\orcidID{0000-0001-9851-1404} \and
Efi Papatheocharous\inst{1}\orcidID{0000-0002-5157-8131} \and
Thomas Olsson\inst{1}\orcidID{0000-0002-2933-1925}}
\authorrunning{Linåker et al.}
%
\institute{RISE Research Institutes of Sweden, Scheelevägen 17, 223 70 Lund, Sweden\\
\email{\{johan.linaker,efi.papatheocharous,thomas.olsson\}ri.se}}
\maketitle              
\begin{abstract}
\textbf{Motivation}: Society's dependence on Open Source Software (OSS) and the communities that maintain the OSS is ever-growing. So are the potential risks of, e.g., vulnerabilities being introduced in projects not actively maintained. By assessing an OSS project's capability to stay viable and maintained over time without interruption or weakening, i.e., the OSS health, users can consider the risk implied by using the OSS as is, and if necessary, decide whether to help improve the health or choose another option. However, such assessment is complex as OSS health covers a wide range of sub-topics, and existing support is limited.
\textbf{Aim}: We aim to create an overview of characteristics that affect the health of an OSS project and enable the assessment thereof.
\textbf{Method}: We conduct a snowball literature review based on a start set of 9 papers, and identify 146 relevant papers over two iterations of forward and backward snowballing. Health characteristics are elicited and coded using structured and axial coding into a framework structure.
\textbf{Results}: 
The final framework consists of 104 health characteristics divided among 15 themes. Characteristics address the socio-technical spectrum of the community of actors maintaining the OSS project, the software and other deliverables being maintained, and the orchestration facilitating the maintenance. Characteristics are further divided based on the level of abstraction they address, i.e., the OSS project-level specifically, or the project's overarching ecosystem of related OSS projects.
\textbf{Conclusion}: 
The framework provides an overview of the wide span of health characteristics that may need to be considered when evaluating OSS health and can serve as a foundation both for research and practice. 

\end{abstract}


\section{Introduction}
Open Source Software (OSS) makes up a pivotal building block in today's digital infrastructure, both in industry and society at large~\cite{blind2021impact}. Due to the nature of OSS, organizations thereby to a large extent become reliant on the external maintenance of the different OSS projects carried out within their respective communities~\cite{eghbal2020working}. If an OSS project is not actively maintained, the risk of vulnerabilities being introduced (either intentionally or not) may rise~\cite{goggins2021open}. These can in turn spread~\cite{ohm2020backstabber} with costly consequences within and between organizations, potentially causing harm to the whole business ecosystems and society in general (cf. Heartbleed\footnote{\url{https://www.cve.org/CVERecord?id=CVE-2014-0160}} and Log4Shell\footnote{\url{https://www.cve.org/CVERecord?id=CVE-2021-44228}}). Organizations, therefore, need to consider the health of the OSS projects that they use to manage the risk coupled with the usage, or dependence of thereof~\cite{butler2022considerations, spinellis2019select}. 

With OSS health, we consider an \textit{OSS project’s capability to stay viable and maintained over time without interruption or weakening}. A topic that we find complex given the wide variety of sub-topics included on the socio-technical spectrum (e.g., toxicity~\cite{fang2022damn}, sponsorships~\cite{overney2020how}, marketing~\cite{miller2022did}, diversity~\cite{foundjem2021onboarding}, badging~\cite{trockman2018adding}, burnout~\cite{miller2019why}, newcomer barriers~\cite{steinmacher2015systematic}). Further complexity is introduced as OSS projects seldom can be considered in isolation due to complex dependency networks of up- and downstream projects~\cite{valiev2018ecosystem, constantinou2017empirical}, implying that both the focal project and its overarching ecosystem of dependencies need to be considered. Extant research does, however, not provide an overview of this wide area of research that OSS health implies (see e.g.,~\cite{rashid2019systematic, steinmacher2015systematic, syeed2013evolution, franco2017open, alves2017software}), even though community and industry-oriented initiatives are starting to emerge~\cite{goggins2021open}. 
 
In this study, our goal is to address this gap by initializing the development of an assessment framework that can provide a comprehensive overview of the health of an OSS project. Specifically, we seek to answer the question \textit{what characteristics affect the health of an OSS project?}

To address the question, we conducted a snowball literature review~\cite{wohlin2014guidelines} with a start set of 9 highly cited papers~\cite{rashid2019systematic, steinmacher2015systematic, valiev2018ecosystem, syeed2013evolution, da2017has, jansen2014measuring, franco2017open, gamalielsson2014sustainability, alves2017software} after which 93 papers were identified in the first iteration, and 53 additional in a second. We structurally coded~\cite{saldana2021coding} the characteristics, related metrics, and underpinning purpose for a paper’s analysis based on an a-priori framework derived from the ecosystem health literature~\cite{manikas2013reviewing, jansen2014measuring}. Through axial coding, we then designed a framework consisting of 104 health characteristics divided among 15 themes.

The assessment framework provides a knowledge base and overview regarding the wide span of health characteristics that may need to be considered when evaluating the health of an OSS project from a socio-technical perspective, including its ecosystem of related OSS projects. The work furthermore provides a foundation for future research by extending and merging prior work in software ecosystem health, as well as that on general OSS health and sustainability topics. For practitioners, the framework provides a reference to establishing an analysis process of OSS projects in use, or in consideration.

\section{Related Work}
Models and approaches have been developed by extant research to evaluate OSS projects from different perspectives, such as sourcing~\cite{badampudi2016software, li2022exploring, spinellis2019select}, alliance partnerships~\cite{shaikh2019selecting}, quality~\cite{yilmazquality, adewumi2013review}, and software ecosystem health~\cite{jansen2014measuring}. These approaches often have a specific focus such as on source code quality~\cite{Gezici2019}, internal capabilities to consume the OSS~\cite{badampudi2016software}, or the potential to build professional business relations through the community~\cite{shaikh2019selecting}. In contrast, there is still a lack of a holistic view of the wider area of research that OSS health implies (see e.g.,~\cite{rashid2019systematic, steinmacher2015systematic, syeed2013evolution, franco2017open, alves2017software}), as well as processes that can help organizations to evaluate OSS projects. 

Interest has, however, emerged around the topic, both in academia~\footnote{\url{https://soheal.github.io/}}, industry~\cite{goggins2021open}, and community settings~\footnote{\url{https://sustainoss.org/}}. Of these, one notable exception exists in the case of the CHAOSS project~\cite{goggins2021open}, a community project collaboratively developing health metrics. These metrics are developed under a set of overarching themes such as value, risk, or evolution. Within each theme, metrics are then provided under several focus areas, each with a specific goal that the metrics aim to provide answers to, e.g., to \textit{``[l]earn about the types and frequency of activities involved in developing code''} under the focus area Code Development Activity within the Evolution theme\footnote{\url{https://github.com/chaoss/wg-evolution}}. No instructions are provided on which metrics to choose. This is instead up to each user to decide.

While the output and intent may overlap, this study differs compared to the CHAOSS project in that it solely considers published research on the topic. Furthermore, our goal is to continue to develop the assessment framework to provide hands-on guidance on what characteristics (and related metrics) to choose, and how these can be operationalized.

\section{Research Design}



We conducted a literature review to answer the research question on OSS health characterization using a snowballing search strategy~\cite{wohlin2014guidelines}. Using a start set of papers, snowballing activity is conducted backward and forwards in iterations. In the backward snowball search, references from a previously included paper in the start set are each reviewed by their title, and place and context in the paper where the reference is used. 
In a forward snowball search, Google Scholar is used to identify papers citing a previously included paper. Papers are reviewed by their title, followed by their abstracts, and a full read, incrementally until a decision can be made on whether the paper should be included or not.


Typically, the pattern continues until no new papers are identified~\cite{wohlin2014guidelines}. In this study, however, the goal is not to conduct a systematic literature review. Rather, we aim to collect a comprehensive knowledge base that can provide an understanding of what characteristics affect the health of an OSS project (RQ), and serve as a foundation for an initial design of an artifact that, iteratively, can be validated and improved in future work through empirical research. Hence, we favor the saturation of characteristics rather than new papers.
Below, we describe our research design and process in further detail. 

\subsection{Inclusion and Exclusion Criteria}
In line with our definition of OSS health, along with guidance from related work~\cite{manikas2013software, jansen2014measuring}, we defined the following inclusion and exclusion criteria (denoted IC and EC respectively):

\begin{small}
\begin{itemize}
    \item[IC] Papers on the growth, attraction, or retention of contributors to OSS projects.
    \item[IC] Papers on the maintainers' ability to maintain their OSS projects.
    \item[IC] Papers on the evaluation of quality or risks of an OSS project related to its health.
    \item[IC] Papers explicitly mentioning the keywords~\cite{manikas2013software, jansen2014measuring} ``health'', ``sustain*'', ``propensity'', ``longevity'', or ``survival'' in the context of OSS in the title, abstract, introduction or conclusion section.
    \item[EC] Not explicitly focused on OSS or focused on (potential) contributors' motivations to engage in an OSS project.
    \item[EC] Published before 2012 and not available in full text.
    \item[EC] Grey and white literature, including book chapters, reports, and student theses.
    \item[EC] Idea and opinion papers, extended abstracts, duplicate studies, secondary studies.
    \item[EC] For extension works most recent published work is included, and others excluded.
    \item[EC] Papers published via other outlets than IEEE, ACM, Springer, Elsevier, or Wiley.
    \item[EC] Non-English papers.
\end{itemize}
\end{small}

\begin{figure}[t]
    \centering
    \includegraphics[width=\columnwidth]{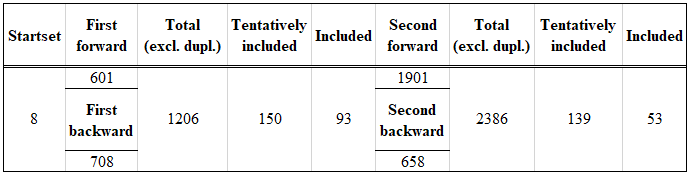}
    \caption{Overview of the number of papers identified and finally included in the first and second iteration of forward and backward snowball searches. Numbers presented in the Total columns are excluding duplicates.}
    \label{fig:review}
\end{figure}

\subsection{Start set}
As a first step, a start set of papers was to be identified as a baseline for the snowballing iterations. Based on keyword searches using Google Scholar it was found that ''health'', ''sustainability'', ''sustain'' and ''survival'' were used interchangeably in alignment with our previously defined definition of OSS health. A search string was constructed accordingly and contextualized to reflect both general and ''ecosystem''-focused literature:

 \textit{(''open source'' OR ''open-source'') AND (''project'' OR ''ecosystem'') AND (''health'' OR ''sustainability'' OR ''sustain'' OR ''survival'')}

This search string was applied through Google Scholar to gain non-biased recommendations to any specific publishing venue~\cite{wohlin2014guidelines}. From the search, three studies were identified as highly cited and seminal papers in the area~\cite{gamalielsson2014sustainability, jansen2014measuring, valiev2018ecosystem} among the first 20 results presented, all passing defined criteria. 

Due to the rather wide definition of OSS health, which is further emphasized by Manikas and Hansen~\cite{manikas2013software}, we decided to specifically look for systematic reviews that may reflect different aspects or areas of the OSS health literature. We, therefore, again using Google Scholar, applied the adapted search string:

\textit{(''open source'' OR ''open-source'') AND (''project'' OR ''ecosystem'') AND (''health'' OR ''sustainability'' OR ''sustain'' OR ''survival'') AND ''systematic'' AND (''review'' OR ''examination'')}

Among the top 20 presented results, five reviews were identified as relevant secondary studies~\cite{franco2017open, alves2017software, steinmacher2015systematic, rashid2019systematic, da2017has}. As they are secondary studies, they are not included in our analysis and only make up a starting point to find primary studies relevant to our study.

\subsection{First Iteration}
In the first iteration, the first author conducted a forward and backward snowball search based on the previously defined start set. The data collection was originally performed in January 2022 but repeated in correlation with the second iteration between the 22nd to 25th of May 2022 to identify recent publications. The (overall) search resulted in an initial set of 150 tentatively included papers from a total of 1206 papers (see Fig.~\ref{fig:review}). 

In the following step, all three authors independently reviewed a sample of 10 percent (i.e., 15 papers) of the tentatively included papers to decide if papers would be included or not. In the same process, the authors also extracted data where applicable. After the independent review process, all three authors discussed and compared findings to arrive at a common understanding of the interpretation and final application of the inclusion and exclusion criteria, as well as the performance of the data extraction process. We continued by separately reviewing and coding papers based on the \textit{a-priori} codes: \textit{Purpose (P)} for analyzing or discussing the health of an OSS project; \textit{Characteristics (C)} that affects or reflects the health of an OSS project; \textit{Metrics (M)} that can quantify, or qualitatively describe, a characteristic.



\begin{figure}[t]
    \centering
    \includegraphics[width=\columnwidth]{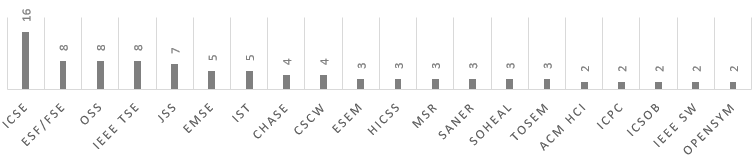}
    \caption{Distribution of studies per venue with two or more.}
    \label{fig:venues}
\end{figure}

During the continued review and coding process, two additional negotiations were performed in terms of agreement of inclusion and coding, each time picking a 10 percent random sample with five papers coded from each of the three authors. This rendered in an overall sample of 30 percent (i.e., 45 papers) being co-analyzed by the three authors. Overall agreement was found on all three occasions with minor adjustments coming out as an effect of the negotiations, e.g., whether a certain extracted element should be considered a purpose, characteristic or metric, or whether a paper should be included or excluded based on the criteria. In the end, 93 papers were included and coded in the first iteration.

Following, we performed a structured coding of the characteristics based on an a-priori-defined health framework inspired by the work of Manikas and Hansen~\cite{manikas2013software} and Jansen~\cite{jansen2014measuring}. The framework consists of two dimensions; the level of abstraction, and the socio-technical dimension.

\begin{small}
\begin{itemize}
    \item Level of abstraction
    \begin{itemize}
        \item \textit{Network-level} concerns characteristics related to the overarching software ecosystem or network that the OSS project is part of, e.g., a language-specific package ecosystem as NPM which in turn consists of multiple OSS components, or the OpenStack ecosystem which in turn consists of numerous of integrating sub-module OSS projects.
        \item \textit{Project-level} concerns characteristics focused explicitly on the OSS project.
    \end{itemize}
    
    \item Socio-technical dimension
    \begin{itemize}
        \item \textit{Actors} concerns the community of developers and users that is part of the OSS project or its overarching software ecosystem.
        \item \textit{Software} concerns the OSS and related artifacts (e.g., documentation) that is being developed by the community of actors, or individuals, that are either part of the OSS project or its overarching ecosystem.
        \item \textit{Orchestration} concerns the governance exercised in terms of development, collaboration, and usage of the software by its community of actors, either within an OSS project or its the overarching ecosystem.
    \end{itemize}
\end{itemize}
\end{small}

After the structural coding, an axial coding process~\cite{saldana2021coding} was performed within each of the six categories of codes that follows by the two dimensions.



\subsection{Second iteration}
\begin{figure}[t!]
    \centering
    \includegraphics[width=\columnwidth]{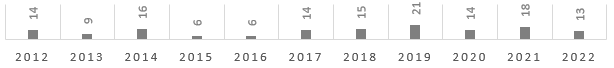}
    \caption{Distribution of studies per year of publication.}
    \label{fig:years}
\end{figure}

In the second iteration, the first author performed the forward snowball, while the third author performed the backward snowball, each resulting in 71 and 68 tentatively included papers out of 1901 and 658 papers respectively. The data collection was performed from the 22nd to the 25th of May 2022. After a second review, 53 papers were included making a total of 146 papers, also considering the first iteration, distributed between 2012-2022 (see Fig.~\ref{fig:years}) over a wide range of venues (see Fig.~\ref{fig:venues}). The 53 papers were coded using the structured and axial coding process as the first iteration. 

In the axial coding, the code book with themes and characteristics generated from the first iteration, structured as per the a-priori framework, was used as a foundation. The coding process rendered in slight modifications, most prominently with the emergence of a new theme focused on security aspects. On the general level, however, we experienced a saturation in the type of characteristics of health that appeared why we decided to not perform any further iterations. 

The final framework was verified through peer debriefing and discussions including all three authors and is further presented in the following section.

\section{Results}

The assessment framework derived from the literature review consists of \textbf{104} health characteristics divided among 15 themes each providing different perspectives on OSS health (see table~\ref{tab:references} in Appendix A). The themes and their underpinning characteristics are structured based on two dimensions: the level of abstraction, and the socio-technical dimension. The level of abstraction considers whether a health characteristic refers to the network- or project-level~\cite{jansen2014measuring, manikas2013reviewing}. The socio-technical dimension considers whether a health characteristic mainly refers to the actors, the software, or the orchestration of the ecosystem~\cite{manikas2013reviewing}. Each characteristic is itself a code incorporating multiple aligning sub-characteristics that have been elicited from identified papers, along with connecting metrics.




Below we provide a summary per theme along with an overarching question contextualizing how the underpinning characteristics affect the health of an OSS project. Codes are provided in parenthesis per characteristic (e.g., a-com-1, meaning [actors-category] - [communication-theme] - and [first characteristic] in the alphabetical order) connects to what papers are related to the characteristic as presented in Table~\ref{tab:references} in Appendix A. Full reference per Paper ID can be found in the online supplementary material to this paper~\cite{SupplementaryMaterial}. We also refer readers to the supplementary material to explore metrics related to the characteristics as identified in the literature, and to investigate the assessment framework in-depth.

\subsection{Actors-oriented characteristics}

\textbf{Communication}:
With communication, we consider the social interactions internally between the actors within an OSS community, and externally by the community in their outward-facing communication. They help to answer the question on \textit{how productive an OSS project is in planning and discussing the evolution and development of its technical and non-technical deliverables}. 

Literature highlights several ways in which the communication takes place, such as mailing lists, issue trackers, and pull-requests. The responsiveness shown by the community through these channels was a commonly referred to characteristic, both considering \textit{response time} (a-com-2), and \textit{response quality} (a-com-2), e.g., in terms of the level of detail, complexity, and correctness. Another characteristic adding to the response-time is the general \textit{social activity} (a-com-3), or frequency, in the communication of the OSS project, e.g., the number of issues opened or comments posed in a certain time interval. Outward-facing communication and \textit{visibility} (a-com-4) was another aspect considered, e.g., how active the community is in terms of social media presence.


\textbf{Culture}:
Cultural characteristics help to answer the question \textit{how able a community is to facilitate a positive and inclusive collaboration and dialogue among existing and potential actors}. They further help to contextualize the social behavior and norms expressed and experienced by the individuals present in a community. The literature emphasizes the experience of contributors, especially in terms of experiencing a personal \textit{satisfaction} (a-cult-2), and being \textit{recognized} for their contributions (a-cult-3), independent of the contribution type and complexity. The presence of \textit{conflicts} (a-cult-1) and how these are managed by a community, as well as the general \textit{openness} (a-cult-5) in terms of the community’s mindset in welcoming and encouraging contributions, inputs, and questions are also highlighted by several studies. The general \textit{sentiment} and tone (a-cult-6) in the communication is another commonly referred to characteristic in literature, where the presence of negative (e.g., insulting, entitled, arrogant, trolling, or unprofessional) and positive qualities (friendliness, welcoming, inclusion) was investigated. \textit{Language heterogeneity} (a-cult-4), or rather the lack of a common language in a community, was a specific concern, investigated by one study, that may cause cross-communication hurdles.


\textbf{Diversity}:
Diversity-related characteristics describe the OSS project's or its overarching ecosystem's ability to be receptive to diversity and self-renew itself, thereby helping to answer the question \textit{how able a community is to accommodate and attract a diverse community of actors, while enabling existing and new use cases of the OSS project}. Among the literature surveyed, multiple angles were covered. One study highlighted the aspect of diversity in the use cases and applications based on the OSS project among its users (a-div-1), aligning with the more general organizational diversity (a-div-3) aspect considering e.g., the size, location, financial stability, business model and influence of the organizations engaged in the community. Demographic diversity (a-div-2), another multifaceted aspect was also investigated, e.g., in terms of gender, culture, and geographical situation. The level and diversity regarding the technical knowledge (a-div-5), e.g., considering programming languages, among individuals was also raised as the diversity among the target users (a-div-4) of the OSS project, e.g., tech savvy users or general end-users.

\textbf{Finance}:
Finance-related characteristics describe the financial support (a-fin-2) in terms of funding and sponsorship provided to or accepted by the OSS community, and the general financial stability (a-fin-1) of the actors in the community that are maintaining or contributing to the OSS project. These characteristics thereby help to answer the question \textit{how financially viable actors are in an OSS community in terms of being able to dedicate their time and resources to the long-term maintenance of the OSS project}.




\textbf{Popularity}:
Characteristics related to popularity describe the general external interest in the OSS project or its overarching ecosystem, helping to answer the question \textit{how popular and well-adopted an OSS project is among existing and potential end-users and contributors}. End-user popularity (a-pop-2), i.e., the level of interest displayed in the project by its consumers, and the external community interest (a-pop-3) shown towards the OSS project were the two most highlighted characteristics in this theme. On an ecosystem-level, one study highlighted the general popularity of the projects hosted within the ecosystem (a-pop-4). The current size of the OSS community in terms of users and developers (a-pop-6), any connection between the OSS project’s use case(s) and the Sustainable Development Goals (a-pop-5), along with the technical inclusion and adoption of an OSS project in downstream software implementations were other characteristics highlighted (a-pop-7). The presence of competing projects was further emphasized as a characteristic that may affect the popularity (a-pop-1). 

\textbf{Stability}:
Characteristics related to stability describe the resilience and robustness of the OSS community or its overarching ecosystem in terms of their population, helping to answer the question \textit{how capable the OSS project is in terms of preserving a critical population of actors with the capability to maintain the OSS project long-term}.
The growth (a-stab-5), retention (a-stab-10), attrition (a-stab-2), and overall turnover (a-stab-12) and size (a-stab-11) of users and developers of an OSS project are characteristics thoroughly investigated by literature. A related characteristic also thoroughly studied is the concentration or distribution of contributions and knowledge to certain individuals or groupings within an OSS project, commonly quantified and described through the bus- or truck factor of a community (a-stab-6). Some studies focused on characterizing OSS projects in terms of their state from a life-cycle perspective (a-stab-7), while others were more forward-looking and focused on predicting future development activity in the OSS project, e.g., in terms of growth or dormancy (a-stab-8).

\textbf{Technical activity}:
The technical activity covers characteristics describing the overall technical activity, helping to answer the question \textit{how productive an OSS project is in evolving and developing its technical and non-technical deliverables}. The theme may, in contrast to the communication theme be considered as the technical pulse of a community and a sign of its productivity, both in terms of technical and non-technical contributions towards the evolution of the concerned OSS project. As per the literature, the technical activity can be considered and evaluated both from the maintainers’ (a-tech-3), contributors’ (a-tech-1), and overall community perspective (a-tech-5). Effectiveness and ease of an OSS project in managing and moving the development forward, e.g., in accepting and reviewing issues and pull-requests, is also highlighted as an important aspect (a-tech-2). Studies also highlight the importance of evaluating the activity in terms of non-code contributions specifically (a-tech-4).

\subsection{Software-oriented characteristics}

\textbf{Development process}:
Characteristics relating to the development process describe the quality and formality of the processes and practices for how the development is performed, addressing  the question \textit{how capable a community is in terms of its development process to maintain the OSS project to a high quality long-term}. The most highlighted characteristic concerns how the onboarding of newcomers to the project is performed, e.g., in terms of mentorship, the introduction of newcomers, and listing of good issues to start with (s-dev-4). A relating characteristic concerns the presence and quality of a contribution process, i.e., how contributions should be made, reviewed, managed, and merged in the OSS project (s-dev-3). Quality and maturity of processes and practices related to quality assurance (s-dev-6), coordination (s-dev-3), coding conventions (s-dev-1), and the development overall (s-dev-5) were also highlighted by different studies. From (primarily) a commercial user perspective, the extent, and quality of any support services provided by the OSS project or the actors engaged in or hosting the project, were also lifted (s-dev-7).

\textbf{Documentation}:
Documentation-related characteristics describe the quality of both general and technical documentation, addressing the question \textit{how capable a community is to develop, persist, and disseminate knowledge among current and future actors engaged in the project}. General documentation encompasses documentation of general nature aimed at both the community, users, and others interested in the OSS project, e.g., readme, homepage, and user manuals (s-doc-5). Technical documentation, on the other hand, refers to documentation covering different aspects of the development process, e.g., in terms of onboarding, planning, contributions, code comments, and quality assurance (s-doc-4). Certain characteristics focus explicitly on quality aspects of the documentation in general, including the completeness (s-doc-1), currentness (s-doc-3), as well as its level of complexity, and ease of understanding (s-doc-2). One study highlighted the availability of the documentation in different languages (s-doc-6).

\textbf{General characteristics}:
A special group of characteristics that help to answer the question on \textit{how attractive an OSS project is based on its general technical features}. These characteristics include general user aspects such the application domain, or product category, of the OSS project (s-gen-2), the type(s) of platforms and operating system(s) that the OSS project is intended for (s-gen-3), its age (s-gen-1), compliance with externally defined standards (s-gen-6), and independence of external software components (s-gen-5). Other characteristics are more technical, such as the OSS project’s size and complexity (s-gen-4), and choice of programming languages, libraries, frameworks, and protocols (s-gen-7).

\textbf{License}:
License-related characteristics were emphasized by several studies, highlighting \textit{how license choices and related practices may affect the popularity and attractiveness of an OSS project, both for commercial actors and individuals}. One study highlights whether there is flexibility in terms of choosing between licenses for the OSS project (s-lic-1). Most studies, however, emphasize the importance of the implications of the license on e.g., redistribution, usage, and packaging (s-lic-2). On a more general level, the quality and presence of practices and processes for license management in the OSS project were emphasized as important for commercial actors (s-lic-4), while the presence of legal jargon was highlighted as a barrier to entry, especially for newcomers (s-lic-3).

\textbf{Scaffolding}:
The scaffolding theme concerns \textit{how robust and accessible the development and communication infrastructure used in the OSS project is in terms of enabling a collaborative and high quality maintenance of the project}. This includes both the availability (s-scaff-5), and accessibility and user-friendliness of tools used for communication and development in the OSS project (s-scaff-4). The presence and quality of continuous integration infrastructure, automation, and practices in the OSS project were also highlighted as important characteristics in terms of software quality and the general attractiveness of a project (s-scaff-2). The ease of setting up the build environment and compiling the OSS project is considered an important aspect to enable newcomers and lower the barrier to adoption of the OSS project (s-scaff-1).

\textbf{Security}:
Characteristics in the security theme help answer the question \textit{how robust an OSS project is in terms of mitigating and managing vulnerabilities and security-related aspects in the current and future maintenance of the project.} More specifically, studies have highlighted past, current, and future (predicted) presence of vulnerabilities and characteristics thereof in dependencies of an OSS project as an important characteristic (s-sec-6). So also the address and persistence of past and current vulnerabilities in an OSS project (s-sec-5). Practices relating to security (s-sec-3), and specifically in terms of dependency management (s-sec-2), e.g., in regards to managing ''conflicting versions of nested dependencies'' as well as updates and security patches, were also raised.

\textbf{Technical quality}:
Technical quality is a rather wide theme considering both the OSS project in general and its code base specifically, helping to answer the question \textit{how robust an OSS project is in terms of its technical quality, considering both a user and developer perspective}. Quality was highlighted both in terms of the product (s-tech-8), component (s-tech-2), architecture (s-tech-1), and source code level (s-tech 9). The complexity of the source code was specifically highlighted in several studies, both in terms of attracting and enabling developers to understand and contribute to the code base, but also in terms of potential correlations to the presence of bugs, vulnerabilities, and negative impact on quality requirements in general. Although several quality requirements were highlighted individually, modularity (s-tech-6), and maintainability (s-tech-5), i.e., the ease of maintaining the source code of the OSS project), were the two that received extra attention.

\subsection{Orchestration-oriented characteristics}

\textbf{Orchestration}:
The orchestration-theme covers characteristics describing the governance structure and quality of the leadership, helping to answer the question of \textit{how mature and open the orchestration is in the OSS project or its overarching ecosystem in terms of enabling an open and inclusive collaboration and long-term maintenance of the OSS project}. Explicitness, formality, and general recognition of the governance structure and leadership were especially highlighted (o-orch-4). As was the way in which the individuals in an OSS community are connected, collaborate, and grouped, explored primarily through the concepts of community patterns and community smells (o-orch-1). The same dimension concerns the overarching ecosystem in how communities collaborate to create resilience and synergies between each other (o-orch-2). Other characteristics regard the leadership's openness to input to decisions and transparency of discussions with actors engaged or with an interest in the OSS project (o-orch-6).

\section{Discussion and Conclusions}


Evaluating the health of an OSS project is a complex exercise. Knowing what to look for, and how to measure it may get out of hand due to a wide focus, or risk becoming too narrow-minded so that important aspects are missed. In this study, we set out to create a comprehensive overview of the wide range of sub-topics related to OSS health. 

Based on a snowball study over two iterations, including 146 primary studies, we derive a framework that consists of 104 health characteristics divided among 15 themes. The themes are dispersed over the socio-technical spectrum with the least coverage in terms of orchestration-related characteristics. It may further be noted that a limited portion of the characteristics is observed on the network level. This relates to the context of the studies included, whether they have focused on an ecosystem (i.e., network) perspective, or the OSS project more specifically. The identified studies confirm, however, the importance of not analyzing an OSS project in isolation. Its dependencies and ties to other projects play an important part, e.g., in terms of resilience and security.

Giving a detailed presentation of the whole framework, including all its characteristics and metrics is beyond the scope and format of this paper, which is why we refer readers to the supplementary material to investigate and explore the assessment framework in-depth~\cite{SupplementaryMaterial}. 

Similar to the CHAOSS project, our framework provides limited guidance in terms of which characteristics to consider, and how. Specifically, we provide limited support in regards to what metrics to operationalize for each characteristic. Readers have to consider metrics as presented through the audit trail and code structure provided in the supplementary material~\cite{SupplementaryMaterial}. In future research, we aim to address this gap through further iterations to design a more mature framework with related processes that can be adopted and tailored based on organizational context and requirements. We aim to leverage case studies, interview surveys and observations of practitioners performing health assessments.

Regarding the limitations in general, it should be noted that we do not claim to have systematically surveyed the literature. Rather, we have made design choices that have limited the search scope and potentially excluded papers (and characteristics) that might be of relevance. We do believe, however, that the snowballing approach has provided a broad sample of the literature, where we could observe a saturation in elicited characteristics in the second iteration. 

A threat to the construct validity regards whether the elicited characteristics are related to the health of an OSS project. To address this concern, we constructed a set of inclusion criteria that we derived from literature, aligning with the research scope of this study. These were then discussed in reoccurring negotiations between the authors throughout the review and coding process to maintain consensus and a common understanding of what characteristics should be elicited, and how these should be coded. By also eliciting the underpinning purpose of each study for the context of why they investigated certain characteristics, we could further judge and characterize the relevance of the study.

\bibliographystyle{splncs04}
\bibliography{references}

\section*{Appendix A}
\label{sec:appendix}

\begin{small}
\begin{longtable}{|p{1cm} p{2.1cm} p{2.8cm} | p{1cm} p{2.1cm} p{2.8cm}|}
\caption{Overview of health characteristics per theme. Characteristics are listed under unique identifiers linking to the respective codes in the online supplementary material~\cite{SupplementaryMaterial}, as is the reference identifiers where P = project level, and N = network-level focus.}
\label{tab:references}\\
\hline
\multicolumn{3}{|l|}{\textbf{Actors / Communication}} & \multicolumn{3}{l|}{\textbf{Actors / Culture}} \\ \hline
a-com-1 & Response-quality & P40, P101, P121, P147 & a-cult-1 & Conflicts & P50, P121, P246 \\
a-com-2 & Response-time & P96, P101, P102, P109,   P121,   P147, P225, P237, P165 & a-cult-2 & Contributor satisfaction & P121, P154 \\
a-com-4 & Social activity & P23, P40, P54, P55, P69,   P121,   P147, P216, P225 & a-cult-3 & Contributor recognition & P78 \\
a-com-5 & Visibility & P87, P104, P153 & a-cult-4 & Language heterogeneity & P121 \\ \cline{1-3}
\multicolumn{3}{|l|}{\textbf{Actors / Diversity}} & a-cult-5 & Openness & P101, P102, P109, P121   P123,   P147. P165 \\ \cline{1-3}
a-div-1 & Application diversity & P78 & a-cult-6 & Sentiment & P27, P43, P101, P114,   P118,   P121, P169, P191, P237 \\ \cline{4-6} 
a-div-2 & Demographic diversity & P75, P87, P104, P162, P172, N94 & \multicolumn{3}{l|}{\textbf{Actors / Finance}} \\ \cline{4-6} 
a-div-3 & Organizational diversity & P13, P78, P116. P122,   P143,   P216, N64, N68, N143 & a-fin-1 & Financial stability & P33, P40, P87, P98, P104 \\
a-div-4 & Target users & P40, P41 & a-fin-2 & Financial support & P40, P82, P98, P148, P208, P260 \\ \cline{4-6} 
a-div-5 & Technical knowledge & P75 & \multicolumn{3}{l|}{\textbf{Actors / Popularity}} \\ \hline
\multicolumn{3}{|l|}{\textbf{Actors / Stability}} & a-pop-1 & Competing projects & P50 \\ \cline{1-3}
a-stab-1 & Age & P77 & a-pop-2 & End-user popularity & P20, P78, P99. P123, P182 \\
a-stab-2 & Attrition & P44, P129 & a-pop-3 & External community interest & P5, P45, P77, P78, P147,   P157, P161, P182, P210, P216, P217, P225, P258 \\
a-stab-3 & Ecosystem growth & N68 & a-pop-4 & Project popularity & N78 \\
a-stab-4 & Forks & N1, N50, N102 & a-pop-5 & SDG & P154 \\
a-stab-5 & Growth & P28, P44, P87, P104, P129 & a-pop-6 & Size & P147 \\
a-stab-6 & Knowledge concentration & P3, P34, P87, P91, P102,   P104,   P163, P164, P171, P173 & a-pop-7 & Technical adoption & P182 \\ \cline{4-6} 
a-stab-7 & Life-cycle stage & P41, P50 & \multicolumn{3}{l|}{\textbf{Actors / Technical activity}} \\ \cline{4-6} 
a-stab-8 & Predicted evolution & P51, P74, P151, P152,   P194,   P202, P216 & a-tech-1 & Contributors' development activity & P85, P121, P147 \\
a-stab-9 & Project growth & N28, N68 & a-tech-2 & Efficiency & P5, P41, P69, P101, P127,   P205,   P222 \\
a-stab-10 & Retention & P15, P44, P47, P49, P96,   P116,   P118, P136, P188, P189, P244, P155 & a-tech-3 & Maintainers' development activity & P69, P102, P131, P147 \\
a-stab-11 & Size & P55, P87, P104, P116,   P208,   P208, P216, N39 & a-tech-4 & Non-code contributions & P227, P234 \\
a-stab-12 & Turnover & P22, P75, P150 & a-tech-5 & Overall development activity & P20, P28, P33, P41, P54,   P69, P77, P78, P80, P85, P87, P98, P99, P104, P109, P115, P116, P123,   P147, P177,   P182, P195, P211, P213, P216, P225, N2, N57,   N64, N68, N94, N143 \\ \hline
\multicolumn{3}{|l|}{\textbf{Software / Development process}} & \multicolumn{3}{l|}{\textbf{Software / Documentation}} \\ \hline
s-dev-1 & Coding conventions & P101, P147, P177 & s-doc-1 & Completeness & P98, P101, P102, P109,   P216,   P165 \\
s-dev-2 & Contribution process & P101, P121, P244 & s-doc-2 & Complexity & P101, P244 \\
s-dev-3 & Coordination & P132 & s-doc-3 & Currentness & P5, P101, P121, P237 \\
s-dev-4 & On-boarding & P35, P46, P101, P102,   P123,   P154, P232, P237, P244, P250 & s-doc-4 & Development docs & P50, P101, P121, P123,   P147,   P165, P237 \\
s-dev-5 & Process maturity & P87, P91, P104, P177, P237 & s-doc-5 & General docs & P50, P123, P147 \\
s-dev-6 & Quality assurance & P5, P50, P98, P101, P123,   P147,   P165 & s-doc-6 & Language availability & P121 \\ \cline{4-6} 
s-dev-7 & Support & P55, P216, P217, P239 & \multicolumn{3}{l|}{\textbf{Software / License}} \\ \hline
\multicolumn{3}{|l|}{\textbf{Software / General factors}} & s-lic-1 & Flexibility & P98 \\ \cline{1-3}
s-gen-1 & Age & P30 & s-lic-2 & Implications & P30, P40, P41, P50, P55,   P109,   P208, P216, P245 \\
s-gen-2 & Application domain & P41 & s-lic-3 & Legal jargon & P121 \\
s-gen-3 & Platform support & P30, P147 & s-lic-4 & Management & P121 \\ \cline{4-6} 
s-gen-4 & Project complexity & P101, P177, P165 & \multicolumn{3}{l|}{\textbf{Software / Scaffolding}} \\ \cline{4-6} 
s-gen-5 & Project independence & P147 & s-scaff-1 & Build environment & P101, P221, P237 \\
s-gen-6 & Standards compliance & P147 & s-scaff-2 & Continuous integration & P50, P165 \\
s-gen-7 & Type of technologies & 30, P50, P121, P147, P165 & s-scaff-3 & Conversation history & P121 \\ \cline{1-3}
\multicolumn{3}{|l|}{\textbf{Software / Security}} & s-scaff-4 & Infrastructure accessibility & P55, P101, P109, P121, P123 \\ \cline{1-3}
s-sec-1 & Dependencies & P98, P101 & s-scaff-5 & Infrastructure availability & P123 \\ \cline{4-6} 
s-sec-2 & Dependency management & P5, P8, P147, P175, P184, P186 & \multicolumn{3}{l|}{\textbf{Orchestration / Orchestration}} \\ \cline{4-6} 
s-sec-3 & Security practices & P29, P147 & o-orch-1 & Community structure & P63, P83, P113, P121, P174,   P176, P198,   P220, P241 \\
s-sec-4 & Trustworthiness & P147, P216 & o-orch-2 & Ecosystem structure & N40, N87, N104 \\
s-sec-5 & Vulnerability persistence & P181, N178, N183 & o-orch-3 & Explicitness of ecosystem & N2 \\
s-sec-6 & Vulnerability presence & P156, P177, N178, N183 & o-orch-4 & Governance & P24, P40, P98, P109,   P147, P208, P216,   P216, N2 \\ \cline{1-3}
\multicolumn{3}{|l|}{\textbf{Software / Technical quality}} & o-orch-5 & Information consistency & N87, N104 \\ \cline{1-3}
s-tech-1 & Architecture quality & P101, P147, P165, N64 & o-orch-6 & KPI-programme & N2 \\
s-tech-2 & Component quality & P33, P60, N33 & o-orch-7 & Openness & P109,   P121 \\
s-tech-3 & Contribution quality & N94 & o-orch-8 & Processes & P9, P50, P102,   P147 \\
s-tech-4 & Ease of integration & P98 & o-orch-9 & Trustworthiness & N87, N104 \\ 
s-tech-5 & Maintainability & P25, P50, P102, P218, P219 & s-tech-8 & Product quality & P33, P40, P99, P104, P109, P217 \\
s-tech-6 & Modularity & P40, P132, P147, P208 & s-tech-9 & Source-code complexity & P66, P77, P98, P101, P121,   P147,   P177, P208 \\
s-tech-7 & Other non-functional requirements & P55, P56, P225, P147 & s-tech-10 & Source-code quality & P62, P101, P109, P121,   P123,   P147, P150, P165, P213 \\

\bottomrule
\end{longtable}
\end{small}

\end{document}